\documentclass[12pt]{article}
\usepackage{ctex}
\usepackage{amssymb,amsmath,color,mathdots}
\usepackage[colorlinks,linkcolor=blue]{hyperref}
\usepackage{geometry}
\usepackage{tabularx}
\usepackage{bm}
\usepackage{authblk}
\usepackage[square, comma, sort&compress, numbers]{natbib}

\title{\bf  The complete weight enumerator of the Reed-Solomon code with dimension two or three}	
\author{\small Canze Zhu}
\author{\small Qunying Liao
	\thanks{Corresponding author.
		
		{~E-mail. qunyingliao@sicnu.edu.cn (Q. Liao), ~canzezhu@163.com (C. Zhu).}	
		
		{~Supported by National Natural Science Foundation of China (Grant No. 12071321).}}
}
\affil[]{\small (College of Mathematical Science, Sichuan Normal University, Chengdu Sichuan, 610066)}
\date{}
	
	\newtheorem{theorem}{Theorem}[section]
	\newtheorem{definition}{Definition}[section]
	\newtheorem{lemma}{Lemma}[section]

	\newtheorem{remark}{Remark}[section]

	\catcode`@=11 \@addtoreset{equation}{section} \catcode`@=12
\begin{document}
	\maketitle
	{\bf Abstract.}
	{\small It is well-known that  Reed-Solomon codes and extended Reed-Solomon codes are two special classes of MDS codes with wide applications in practice.  The complete weight enumerators of these codes are very  important for determining the  capability of both error-detection and error-correction. In this paper, for any positive integer $m$ and prime $p$, basing on the character sums,  we determine  the complete weight enumerators of the Reed-Solomon code and the extended Reed-Solomon code with dimension $k$ $(k=2,3)$ over $\mathbb{F}_{p^m}$, explictly, which are generalizations of the corresponding results in \cite{BK91,K04}.}\\

	{\bf Keywords.}	{\small Reed-Solomon codes; Extended Reed-Solomon codes; Complete weight enumerators; Character sums.}

	\section{Introduction}
For any positive integer $m$ and prime $p$, let $\mathbb{F}_{p^m}$ be the finite field with $p^m$ elements.  An $[n,k,d]$ linear code $\mathcal{C}$ over $\mathbb{F}_{p^m}$ is a $k$-dimensional subspace of $\mathbb{F}_{p^m}^n$ with minimum (Hamming) distance $d$ and length $n$. Let $A_i\ (i=1,\ldots,n)$ be the number of codewords with Hamming weight $i$ in $\mathcal{C}$, then the weight distribution  of $\mathcal{C}$ is defined by  $(1,A_1,\ldots,A_n)$. In addition, the complete weight enumerator of a codeword $\mathbf{c}$ is the monomial
{\small	$$w(\mathbf{c})=\prod_{\rho\in \mathbb{F}_{p^m}}w_\rho^{t_\rho}$$}
	in the variables $w_\rho$ $(\rho\in\mathbb{F}_{p^m})$, where $t_\rho$ $(\rho\in\mathbb{F}_{p^m})$ denotes the number of components of $\mathbf{c}$ equal to $\rho$. The complete weight enumerator of $\mathcal{C}$ is defined to be
	{\small	$$W(\mathcal{C})=\sum\limits_{\mathbf{c}\in\mathcal{C}}w(\mathbf{c}).$$}
	 
	 It is well-known that  the weight distribution can be deduced from the complete weight enumerator, and it is key  to determine the  capability for both error-detection and error-correction \cite{T07}. Furthermore, the complete weight enumerator plays an important role in the investigations of many areas, such as the constructions of optimal constant composition codes \cite{D1}, the deception probability of some authentication codes \cite{D2}, and so on. 
	
	Contributed to the parameters achieving the Singleton bound, the MDS code  can be applied to correct maximal number of errors for given code rate \cite{HP10}. Especially,  Reed-Solomon (in short, RS) codes and extended Reed-Solomon (in short, ERS) codes are two classes of MDS codes  with wide applications in practice.

Naturally, it is interesting to determine the complete weight enumerators of  RS (ERS) codes, while there are few works in this topic.
In 1991, by solving some high-order equations over an extension field of $\mathbb{F}_{2^m}$, Blake and Kith determined the complete weight enumerators of both the primitive RS code and the standard RS code over $\mathbb{F}_{2^m}$ with dimension $k$ $(k=2,3,4)$ \cite{BK91}. In 2004,  by using Blake and Kith's techniques, Toshiyuki derived the complete weight enumerator of the extended standard RS code over $\mathbb{F}_{2^m}$ with dimension $k$ $(k=2,3)$  \cite{K04}. The other aspects for the complete weight enumerators of RS codes over $\mathbb{F}_{2^m}$ can be seen \cite{TT1993,TT1995}. 
	
	In this paper, for any prime $p$,  by employing the character sums, we determine the complete weight enumerator of RS (ERS) codes  over $\mathbb{F}_{p^m}$  with dimension  $k$ $(k=2,3)$,  which are generalizations of the corresponding results in \cite{BK91,K04}.
	
	This paper is organized as follows. In section 2, we give some related basic notations and results of RS (ERS) codes and character sums, respectively. In section 3,  we present our results.	In section 4, we prove the main results. 
\section{Preliminaries}
In this section, we give some notations and results of RS (ERS) codes and character sums, respectively.
\subsection{RS (ERS) codes}
\begin{definition} [\cite{HP10,MS77}]
	For positive integers  $k$ and $n$ with $k\le n\le p^m$, let $\boldsymbol{\alpha}=(\alpha_1,\ldots,\alpha_n)$ be an $n$-tuple of distinct elements in $\mathbb{F}_{p^m}$,
	Then the RS code is  defined as
{\small	\begin{align*}
	\mathcal{RS}_{k}(\boldsymbol{\alpha})=\big\{\big(f({\alpha_{1}}),\ldots,f(\alpha_{n})\big)~|~ f(x)\in \mathbb{F}_{p^m}[x],~\deg f(x)\le k-1\big\},
	\end{align*}	}
	and the ERS code is defined as
{\small	\begin{align*}
	\mathcal{ERS}_{k}(\boldsymbol{\alpha})=\big\{\big(f({\alpha_{1}}),\ldots,f(\alpha_{n}),f_{k-1}\big)~|~ f(x)\in \mathbb{F}_{p^m}[x],~\deg f(x)\le k-1\big\},
	\end{align*}}
	where $f_{k-1}$ is the coefficient of $x^{k-1}$ in $f(x)$.
\end{definition}
\begin{remark}
	Let $\mathbb{F}_{p^m}^{*}=\{\alpha_1,\ldots,\alpha_{p^m-1}\}$. If $\boldsymbol{\alpha}=(\alpha_1,\ldots,\alpha_{p^m-1})$, then $\mathcal{RS}_{k}(\boldsymbol{\alpha})$ is called the primitive RS code. If $\boldsymbol{\alpha}=(0,\alpha_1,\ldots,\alpha_{p^m-1})$, then $\mathcal{RS}_{k}(\boldsymbol{\alpha})$ is called the standard RS code. 
\end{remark}
\subsection{Character sums}
	    \indent An additive character $\chi$ of $\mathbb{F}_{p^m}$ is a function from $\mathbb{F}_{p^m}$ to the multiplicative group $U=\{u\ |\ |u|=1,\ u\in\mathbb{C}\}$, such that $\chi(x+y)=\chi(x)\chi(y)$ for any $x,\ y\in\ \mathbb{F}_{p^m}$. For each $b\in \mathbb{F}_{p^m}$, the function
	\begin{center}
		$\chi_b(x)=\zeta_p^{\mathrm{Tr}(bx)}$ \quad $(x\in \mathbb{F}_{p^m})$
	\end{center}
	defines an additive character over $\mathbb{F}_{p^m}$. The character $\chi:=\chi_1$ is called the canonical additive character of $\mathbb{F}_{p^m}$. The orthogonal property for the additive characters is given by
	\begin{align*}
	\sum_{x\in \mathbb{F}_{p^m}}\zeta_p^{\mathrm{Tr}(bx)}=\begin{cases}
	p^m,\quad& \text{if}~b=0;\\
	0,\quad &\text{otherwise}.
	\end{cases}
	\end{align*}     
	We extend the quadratic character $\eta$ over $\mathbb{F}_{p^m}^*$ by letting $\eta(0)=0$, then the quadratic Gauss sums $G(\eta,\chi)$ over $\mathbb{F}_{p^m}$ is defined as 
	\begin{center}
		$G(\eta,\chi)=\sum\limits_{x\in \mathbb{F}_{p^m}}\eta(x)\chi(x)$.
	\end{center}
	
	\indent     Now, some related results for quadratic Gauss sums  and character sums are given as follows. 
	
		\begin{lemma}[\cite{LN97}, Theorem 5.15]\label{l22}
		For the Gauss sum $G(\eta,\chi)$ over $\mathbb{F}_{p^m}$,
		\begin{align*}
		G(\eta,\chi)=(-1)^{m-1}\sqrt{-1}^\frac{(p-1)^2m}{4}p^{\frac{m}{2}}.
		\end{align*}
	\end{lemma}
	
	\begin{lemma}[\cite{LN97}, Theorem 5.33]\label{l23}
		Let $f(x)=a_2x^2+a_1x+a_0\in \mathbb{F}_{p^m}[x]$ with $a_2\neq0$, then
		\begin{align*}
		\sum_{c\in \mathbb{F}_{p^m}}\chi(f(c))=G(\eta,\chi)\eta(a_2)\chi(a_0-a_1^2(4a_2)^{-1}).
		\end{align*}
	\end{lemma}

\section{Complete Weight Enumerator of the RS (ERS) Code with Dimension $k$ $(k=2,3)$}
\subsection{Main results}
\begin{theorem}\label{c30}
	For any prime $p$ and positive integers $m$ and $n\ge2$, let $\boldsymbol{\alpha}=(\alpha_1,\ldots,\alpha_n)$ be an $n$-tuple of distinct elements in $\mathbb{F}_{p^m}$, then the complete weight enumerators of $\mathcal{RS}_2({\boldsymbol{\alpha}})$ and $\mathcal{ERS}_2({\boldsymbol{\alpha}})$ are given in $(\ref{l31})$-$(\ref{l32})$, respectively,
	\begin{align}\label{l31}
	\mathrm{W}(\mathcal{RS}_2({\boldsymbol{\alpha}}))=\sum_{\rho\in\mathbb{F}_{p^m}}w_{\rho}^{n}+\sum_{\gamma_0\in\mathbb{F}_{p^m}}\sum_{\gamma_1\in\mathbb{F}_{p^m}^*}\prod_{\rho\in\{\gamma_0+\gamma_1\alpha_i|i=1,\ldots,n\}}w_\rho,
	\end{align}
	\begin{align}\label{l32}
	\mathrm{W}(\mathcal{ERS}_2({\boldsymbol{\alpha}}))=\sum_{\rho\in\mathbb{F}_{p^m}}w_{\rho}^{n}+\sum_{\gamma_0\in\mathbb{F}_{p^m}}\sum_{\gamma_1\in\mathbb{F}_{p^m}^*}w_{\gamma_{1}}\prod_{\rho\in\{\gamma_0+\gamma_1\alpha_i|i=1,\ldots,n\}}w_\rho.
	\end{align}
\end{theorem}
\begin{remark}
	Let $\mathbb{F}_{p^m}^{*}=\{\alpha_1,\ldots,\alpha_{p^m-1}\}$.
	
	 (1) By taking $\boldsymbol{\alpha}=(\alpha_1,\ldots,\alpha_{p^m-1})$ or  $(0,\alpha_1,\ldots,\alpha_{p^m-1})$  in $(\ref{l31})$, respectively, one can obtain the results of section 3 in \cite{BK91}. 
	 
	 (2) By taking $\boldsymbol{\alpha}=(0,\alpha_1,\ldots,\alpha_{p^m-1})$ in $(\ref{l32})$, one can obtain Theorem $1$ in \cite{K04}.
\end{remark}
\begin{theorem}\label{c31}
	For any positive integer $m$ and prime $p$ with $p^m\ge 3$, let $\mathbb{F}_{p^m}=\{\alpha_1,\ldots,\alpha_{p^m}\}$ and ${\boldsymbol{\alpha}}=(\alpha_1,\ldots,\alpha_{p^m})$, then the following two asserctions hold.\\	
	
	$(1)$ For $p=2$, we have
	{\small \begin{align}\label{c311}
		\mathrm{W}(\mathcal{RS}_3(\boldsymbol{\alpha}))\! 
		=\!\!\sum_{\rho\in\mathbb{F}_{2^m}}\!w_\rho^{2^{m}}\!+(2^m-1)2^{m+1}\!\!\prod_{\rho\in \mathbb{F}_{2^m}}\!w_{\rho}\!	+(2^m-1)\sum_{\gamma_1\in\mathbb{F}_{2^m}^*}\!\!\sum_{\gamma_0\in\mathbb{F}_{2^m}}\!\!\prod_{\substack{\rho\in\mathbb{F}_{2^m}\\\mathrm{Tr}(\rho)=0}}w_{\gamma_1\rho+\gamma_0}^2,
		\end{align}}
	and
	{\begin{align}\label{c312}\begin{aligned}	
		\mathrm{W}(\mathcal{ERS}_3(\boldsymbol{\alpha}))
		=&\sum_{\rho\in\mathbb{F}_{2^m}}w_0w_\rho^{2^{m}}+(2^m-1)2^mw_0\prod_{\rho\in \mathbb{F}_{2^m}}w_{\rho}
		+2^m\sum_{\gamma_2\in\mathbb{F}_{p^m}^*}w_{\gamma_2} \prod_{\rho\in\mathbb{F}_{p^m}}w_{\rho}	\\&+\sum_{\gamma_2\in\mathbb{F}_{2^m}^*}\sum_{\gamma_1\in\mathbb{F}_{2^m}^*}\sum_{\gamma_0\in\mathbb{F}_{2^m}}w_{\gamma_2}\prod_{\substack{\rho\in\mathbb{F}_{2^m}\\\mathrm{Tr}(\rho)=0}}w_{\gamma_1\rho+\gamma_0}^2.
		\end{aligned}
		\end{align}}\\
	
	$(2)$ For odd $p$, we have 
	{\small \begin{align}\label{c314}\begin{aligned}	
		W(\mathcal{RS}_3(\boldsymbol{\alpha}))
		=&	\sum_{\rho\in\mathbb{F}_{p^m}}w_\rho^{p^{m}}+(p^m-1)p^m\prod_{\rho\in \mathbb{F}_{p^m}}w_{\rho}
		\\&+\frac{(p^m-1)p^m}{2}\sum_{\epsilon\in\{-1,1\}}\sum_{\gamma_1\in\mathbb{F}_{p^m}}w_{\gamma_1}\prod_{\rho\in\mathbb{F}_{p^m}\backslash\{\gamma_1\}}
		w_{\rho}^{1+\epsilon\eta(\rho-\gamma_1)},
		\end{aligned}
		\end{align}}
	and
	{\small \begin{align}\label{c315}\begin{aligned}	
		W(\mathcal{ERS}_3(\boldsymbol{\alpha}))=&p^m	\sum_{\rho\in\mathbb{F}_{p^m}}w_0w_\rho^{p^{m}}+(p^m-1)p^mw_0\prod_{\rho\in \mathbb{F}_{p^m}}w_{\rho}\\&+p^m\sum_{\gamma_2\in\mathbb{F}_{p^m}^*}\sum_{\gamma_1\in\mathbb{F}_{p^m}} w_{\gamma_2} w_{\gamma_1}\prod_{\rho\in\mathbb{F}_{p^m}\backslash\{\gamma_1\}}
		w_{\rho}^{1+\eta(\gamma_2)\eta(\rho-\gamma_1)}.
		\end{aligned}
		\end{align}}
\end{theorem}
\begin{theorem}\label{c32}
	For any positive integer $m$ and prime $p$ with $p^m\ge 4$, let $\beta\in\mathbb{F}_{p^m}$, $\mathbb{F}_{p^m}\backslash \{\beta\}=\{\alpha_1,\ldots,\alpha_{p^m-1}\}$ and ${\boldsymbol{\alpha}}=(\alpha_1,\ldots,\alpha_{p^m-1})$, then the following two asserctions hold.\\
	
	$(1)$ For $p=2$, we have
	{\small \begin{align}\label{c321}
		\begin{aligned}
		\mathrm{W}(\mathcal{RS}_3(\boldsymbol{\alpha}))=\sum_{\rho\in\mathbb{F}_{2^m}}w_\rho^{2^{m}}+2(2^m-1)\prod_{\rho\in \mathbb{F}_{2^m}}w_{\rho}	+(2^m-1)\sum_{\gamma_1\in\mathbb{F}_{2^m}^*}\sum_{\gamma_0\in\mathbb{F}_{2^m}}w_{\gamma_{0}}\prod_{\substack{\rho\in\mathbb{F}_{2^m}^{*}\\\mathrm{Tr}(\rho)=0}}w_{\gamma_1\rho+\gamma_0}^2,
		\end{aligned}
		\end{align}}and
	{\small \begin{align}\label{c322}\begin{aligned}	
		\mathrm{W}(\mathcal{ERS}_3(\boldsymbol{\alpha}))
		=&\sum_{\rho\in\mathbb{F}_{2^m}}w_0w_\rho^{2^{m}-1}+(2^m-1)\sum_{\gamma_1\in\mathbb{F}_{2^m}}w_0\prod_{\rho\in \mathbb{F}_{2^m}\backslash\{\gamma_1\}}w_{\rho}
		\\&+\sum_{\gamma_2\in\mathbb{F}_{p^m}^*}\sum_{\gamma_1\in\mathbb{F}_{2^m}}w_{\gamma_2}\prod_{\rho\in\mathbb{F}_{2^m}\backslash\{\gamma_1\}}w_{\rho}	+\sum_{\gamma_2\in\mathbb{F}_{2^m}^*}\sum_{\gamma_1\in\mathbb{F}_{2^m}^*}\sum_{\gamma_0\in\mathbb{F}_{2^m}}w_{\gamma_2}w_{\gamma_0}\prod_{\substack{\rho\in\mathbb{F}_{2^m}^{*}\\\mathrm{Tr}(\rho)=0}}w_{\gamma_{1}\rho+\gamma_0}^2.
		\end{aligned}
		\end{align}}

	$(2)$ For odd $p$, we have
	{\small \begin{align}\label{c324}\begin{aligned}	
		\mathrm{W}(\mathcal{RS}_3(\boldsymbol{\alpha}))
		=&	\sum_{\rho\in\mathbb{F}_{p^m}}w_\rho^{p^m-1}+(p^m-1)\sum_{\gamma\in\mathbb{F}_{p^m}}\prod_{\rho\in \mathbb{F}_{p^m}\backslash\{\gamma\}}w_{\rho}\\
		&+\frac{p^m-1}{2}\sum_{\epsilon\in\{-1,1\}}\sum_{\gamma_1\in\mathbb{F}_{p^m}}\prod_{\rho\in\mathbb{F}_{p^m}\backslash\{\gamma_1\}}
		w_{\rho}^{1+\epsilon\eta(\rho-\gamma_1)}\\
		&+(p^m-1)\sum_{\epsilon\in\{-1,1\}}\sum_{\gamma_1\in\mathbb{F}_{p^m}}\sum_{\substack{\gamma_0\in\mathbb{F}_{p^m}\\\eta(\gamma_0)=\epsilon}}w_{\gamma_1}w_{\gamma_0+\gamma_{1}}\prod_{\rho\in\mathbb{F}_{p^m}\backslash\{\gamma_1,\gamma_0+\gamma_1\}}w_{\rho}^{1+\epsilon\eta(\rho-\gamma_1)},
		\end{aligned}
		\end{align}}and
	{\small \begin{align}\label{c325}\begin{aligned}	
		\mathrm{W}(\mathcal{ERS}_3(\boldsymbol{\alpha}))
		=&w_0	\sum_{\rho\in\mathbb{F}_{p^m}}w_\rho^{p^m-1}+(p^m-1)\sum_{\gamma_0\in\mathbb{F}_{p^m}}w_0\prod_{\rho\in \mathbb{F}_{p^m}\backslash\{\gamma_0\}}w_{\rho}
		\\&+\sum_{\gamma_2\in\mathbb{F}_{p^m}^{*}}\sum_{\gamma_1\in\mathbb{F}_{p^m}}w_{\gamma_2}\prod_{\rho\in\mathbb{F}_{p^m}\backslash\{\gamma_1\}}
		w_{\rho}^{1+\eta(\gamma_2)\eta(\rho-\gamma_1)}\\
		&+2\sum_{\epsilon\in\{-1,1\}}\sum_{\substack{\gamma_2\in\mathbb{F}_{p^m}^{*}\\\eta(\gamma_2)=\epsilon}}\sum_{\gamma_1\in\mathbb{F}_{p^m}}\sum_{\substack{\gamma_0\in\mathbb{F}_{p^m}\\\eta(\gamma_0)=\epsilon}}w_{\gamma_2}w_{\gamma_1}w_{\gamma_0+\gamma_{1}}\prod_{\rho\in\mathbb{F}_{p^m}\backslash\{\gamma_1,\gamma_0+\gamma_1\}}w_{\rho}^{1+\epsilon\eta(\rho-\gamma_1)}.\\
		\end{aligned}
		\end{align}}
\end{theorem}

\begin{remark}

(1) By taking $\beta=0$ in Theorem $\ref{c32}$ $(\ref{c321})$ and Theorem $\ref{c31}$ $(\ref{c311})$, we can get the corresponding results in section 4 \cite{BK91}. 

 (2) By taking $\beta=0$ in Theorem $\ref{c32}$ $(\ref{c322})$, we can get Theorem $2$ in \cite{K04}.
\end{remark}

\section{Proofs for Main Results}
\subsection{Some auxiliary lemmas}
In this subsection, we give some necessary lemmas to prove Theorems \ref{c30}-\ref{c32}. 

For any positive integers $k$ and $n$ with $k\le n$, let  $\boldsymbol{\alpha}=(\alpha_1,\ldots,\alpha_n)$ be an $n$-tuple of distinct elements in $\mathbb{F}_{p^m}$, then the complete weight enumerators of  $\mathcal{RS}_k(\boldsymbol{\alpha})$ and $\mathcal{ERS}_k(\boldsymbol{\alpha})$ are both
 depended on the value distribution of $|{N}_{D}(a_{k-1},\ldots,a_{0},\rho)|$, where $${N}_{D}(a_{k-1},\ldots,a_1,a_{0},\rho)=\big\{x\in D~|~a_{k-1}x^{k-1}+\cdots+a_1x+a_0=\rho\big\}$$ with
$\rho, a_{0},\ldots,a_{k-1}\in\mathbb{F}_{p^m}$ and  $D=\{\alpha_1,\ldots,\alpha_n\}$.
Furthermore, by the orthogonal proprety for additive characters, we have
\begin{align}\label{NDS}
\begin{aligned}
|{N}_{D}(a_{k-1},\ldots,a_{0},\rho)|=&\sum_{x\in D}\bigg(\frac{1}{p^m}\sum_{z\in\mathbb{F}_{p^m}}\zeta_p^{\mathrm{Tr}(z(a_{k-1}x^{k-1}+\cdots+a_1x+a_0-\rho))}\bigg)\\
=&\frac{1}{p^m}\sum_{z\in\mathbb{F}_{p^m}}\sum_{x\in D}\zeta_p^{\mathrm{Tr}(z(a_{k-1}x^{k-1}+\cdots+a_1x+a_0-\rho))}.
\end{aligned}
\end{align}
Especially, for $k=3$, $\beta\in\mathbb{F}_{p^m}$ and $D=\mathbb{F}_{p^m}\text{~or~}\mathbb{F}_{p^m}\backslash\{\beta\}$, the explicit value for $|{N}_{D}(a_{2},a_{1},a_{0},\rho)|$ is given in Lemmas \ref{lm3qw}-\ref{lm3qw1}.
\begin{lemma}\label{lm3qw}
	For any positive integer $m$  and prime $p$, we have\\
	\begin{align*}
	|N_{\mathbb{F}_{p^m}}(0,a_1,a_0,\rho)|=\begin{cases}
	p^m,\quad&\text{~if~}a_1=0\text{~and~}\rho=a_0;\\
	0,\quad&\text{~if~} a_1=0\text{~and~}\rho\neq a_0;\\
	1,\quad &\text{~if~}a_1\neq 0.
	\end{cases}
	\end{align*}
And for $a_2\neq 0$, \\	
	
	$(1)$ if $p=2$, then
	\begin{align}\label{l1}
	\begin{aligned}
	|N_{\mathbb{F}_{p^m}}(a_2,a_1,a_0,\rho)|=\begin{cases}
	1,\quad &\text{~if~}a_1=0;\\	
	2,\quad &\text{~if~}a_1\neq 0\text{~and~}\mathrm{Tr}(a_2a_1^{-2}(\rho-a_0))=0;\\
	0,\quad &\text{~if~}a_1\neq 0\text{~and~}\mathrm{Tr}(a_2a_1^{-2}(\rho-a_0))=1;
	\end{cases}
	\end{aligned}
	\end{align}	
	
	$(2)$ if $p$ is odd, then{\small
	\begin{align}\label{l2}
	|N_{\mathbb{F}_{p^m}}(a_2,a_1,a_0,\rho)|=\begin{cases}
	1,&\!\!\!\!\text{if~}\rho=(4a_2)^{-1}(4a_0a_2-a_1^2);\\
	1+\eta(a_2)\eta\left(\rho-(4a_2)^{-1}(4a_0a_2-a_1^2)\right),&\!\!\!\!\text{if~}\rho\neq (4a_2)^{-1}(4a_0a_2-a_1^2).
	\end{cases}
	\end{align}}
	
\end{lemma}

{\bf Proof.}~By taking $k=3$ and $D=\mathbb{F}_{p^m}$ in $(\ref{NDS})$, we have

\begin{align}\label{NDS3}
\begin{aligned}
|N_{\mathbb{F}_{p^m}}(a_2,a_1,a_0,\rho)|=&\frac{1}{p^m}\sum_{x\in\mathbb{F}_{p^m}}\sum_{z\in\mathbb{F}_{p^m}}\zeta_p^{\mathrm{Tr}\big(z(a_{2}x^2+a_1x+a_0-\rho)\big)}\\
=&1+\frac{1}{p^m}\sum_{z\in\mathbb{F}_{p^m}^{*}}\zeta_p^{\mathrm{Tr}((a_0-\rho)z)}\sum_{x\in\mathbb{F}_{p^m}}\zeta_p^{\mathrm{Tr}\big(a_2zx^2+a_1zx\big)}\\
=&1+\frac{1}{p^m}\Omega,
\end{aligned}
\end{align}
where
\begin{align*}
\Omega=\sum_{z\in\mathbb{F}_{p^m}^{*}}\zeta_p^{\mathrm{Tr}\big((a_0-\rho)z\big)}\sum_{x\in\mathbb{F}_{p^m}}\zeta_p^{\mathrm{Tr}\big(a_2zx^2+a_1zx\big)}.
\end{align*}

Therefore, for $a_2=0$, we have 
\begin{align}\label{NDS31}
\begin{aligned}
\Omega&=\sum_{z\in\mathbb{F}_{p^m}^{*}}\zeta_p^{\mathrm{Tr}((a_0-\rho)z)}\sum_{x\in\mathbb{F}_{p^m}}\zeta_p^{\mathrm{Tr}(a_1zx)}\\
&=\begin{cases}
p^m\sum\limits_{z\in\mathbb{F}_{p^m}^{*}}\zeta_p^{\mathrm{Tr}((a_0-\rho)z)},\quad &\text{~if~}a_1=0;\\	
0,\quad &\text{~if~}a_1\neq 0,\\
\end{cases}\\
&=\begin{cases}
(p^m-1)p^m, &\text{~if~}a_1=0\text{~and~}a_0=\rho;\\	
-p^m, &\text{~if~}a_1=0\text{~and~}a_0\neq\rho;\\	
0,\quad &\text{~if~}a_1\neq 0.\\
\end{cases}
\end{aligned}
\end{align}\\

Now for $a_2\neq 0$,  we have the following two cases.

$(1)$ If $p=2$, then
\begin{align}\label{NDS3e}
\begin{aligned}
\Omega&=\sum_{z\in\mathbb{F}_{2^m}^{*}}\zeta_p^{\mathrm{Tr}((a_0-\rho)z)}\sum_{x\in\mathbb{F}_{2^m}}\zeta_p^{\mathrm{Tr}\big(\big((a_2z)^{2^{m-1}}x\big)^2+a_1zx\big)}\\
&=\sum_{z\in\mathbb{F}_{2^m}^{*}}\zeta_p^{\mathrm{Tr}((a_0-\rho)z)}\sum_{x\in\mathbb{F}_{2^m}}\zeta_p^{\mathrm{Tr}\big(\big((a_2z)^{2^{m-1}}+a_1z\big)x\big)}\\
&=2^m\sum_{\substack{z\in\mathbb{F}_{2^m}^{*}\\(a_2z)^{2^{m-1}}+a_1z=0}}\zeta_p^{\mathrm{Tr}((a_0-\rho)z)}\\
&=\begin{cases}
0,\quad &\text{~if~}a_1=0;\\	
2^m\zeta_p^{\mathrm{Tr}(a_2a_1^{-2}(\rho-a_0))},\quad &\text{~if~}a_1\neq 0,
\end{cases}\\
&=\begin{cases}
0,\quad &\text{~if~}a_1=0;\\	
2^m,\quad &\text{~if~}a_1\neq 0\text{~and~}\mathrm{Tr}(a_2a_1^{-2}(\rho-a_0))=0;\\
-2^m,\quad &\text{~if~}a_1\neq 0\text{~and~}\mathrm{Tr}(a_2a_1^{-2}(\rho-a_0))=1.
\end{cases}
\end{aligned}
\end{align}

$(2)$ If $p$ is odd, then by Lemmas \ref{l22}-\ref{l23}, we have
\begin{align}\label{NDS3o}
\begin{aligned}
\Omega&=\eta(a_2)G(\eta,\chi)\sum_{z\in\mathbb{F}_{p^m}^{*}}\eta(z)\zeta_p^{\mathrm{Tr}\big((4a_2)^{-1}((4a_0a_2-a_1^2)-4a_2\rho) z\big)}\\
&=\begin{cases}
0,\quad &\text{~if~}\rho=(4a_2)^{-1}(4a_0a_2-a_1^2);\\	
\eta(a_2)\eta\left((4a_2)^{-1}(4a_0a_2-a_1^2)-\rho\right)G(\eta,\chi)^2,\quad &\text{~if~}\rho\neq (4a_2)^{-1}(4a_0a_2-a_1^2),\\
\end{cases}\\
&=\begin{cases}
0,\quad &\text{~if~}\rho=(4a_2)^{-1}(4a_0a_2-a_1^2);\\	
\eta(a_2)\eta\left(\rho-(4a_2)^{-1}(4a_0a_2-a_1^2)\right)p^m,\quad &\text{~if~}\rho\neq (4a_2)^{-1}(4a_0a_2-a_1^2).
\end{cases}
\end{aligned}
\end{align}

Now by $(\ref{NDS3})$-$(\ref{NDS3o})$,  we complete the proof.
$\hfill\Box$\\

\begin{lemma}\label{lm3qw1}	For any positive integer $m$, prime $p$ and $\beta\in\mathbb{F}_{p^m}$, we have
	\begin{align*}
	|N_{\mathbb{F}_{p^m}\backslash\{\beta\}}(0,a_1,a_0,\rho)|=\begin{cases}
	p^m-1,\quad&\text{~if~}a_1=0\text{~and~}\rho=a_0;\\
	0,\quad&\text{~if~} a_1=0\text{~and~}\rho\neq a_0; \\
	0,&\text{~if~} a_1\neq 0\text{~and~}\rho=a_1\beta+a_0;\\
	1,\quad &\text{~if~}a_1\neq 0\text{~and~}\rho\neq a_1\beta+a_0.
	\end{cases}
	\end{align*}
And for $a_2\neq 0$, 	

	$(1)$ 	if $p=2$, then
	\begin{align}\label{23}
	\begin{aligned}
	&|N_{\mathbb{F}_{2^m}\backslash\{\beta\}}(a_2,a_1,a_0,\rho)|\\=&\begin{cases}
	0,\quad &\text{~if~}a_1=0\text{~and~}\rho= a_2\beta^2+a_0;\\	
	1,\quad &\text{~if~}a_1=0\text{~and~}\rho\neq a_2\beta^2+a_0;\\	
	1,\quad &\text{~if~}a_1\neq 0\text{~and~}\rho= a_2\beta^2+a_1\beta+a_0;\\
	2,\quad &\text{~if~}a_1\neq 0,\mathrm{Tr}(a_2a_1^{-2}(\rho-a_0))=0\text{~and~}\rho\neq a_2\beta^2+a_1\beta+a_0;\\
	0,\quad &\text{~if~}a_1\neq 0\text{~and~}\mathrm{Tr}(a_2a_1^{-2}(\rho-a_0))=1;
	\end{cases}
	\end{aligned}
	\end{align}\\

	$(2)$ if $p$ is odd, then
	{\small
		\begin{align}\label{p3}\begin{aligned}
		&|N_{\mathbb{F}_{p^m}\backslash\{\beta\}}(a_2,a_1,a_0,\rho)|\\
		=&\begin{cases}
		0,&\!\!\text{if~}\rho=(4a_2)^{-1}(4a_0a_2-a_1^2)= a_2\beta^2+a_1\beta+a_0;\\
		1,&\!\!\text{if~}\rho=(4a_2)^{-1}(4a_0a_2-a_1^2)\text{~and~}\rho\neq a_2\beta^2+a_1\beta+a_0;\\
		\eta(a_2)\eta(\rho-(4a_2)^{-1}(4a_0a_2-a_1^2)),&\!\!\text{if~}\rho\neq (4a_2)^{-1}(4a_0a_2-a_1^2)\text{~and~}\rho= a_2\beta^2+a_1\beta+a_0;\\
		1+\eta(a_2)\eta(\rho-(4a_2)^{-1}(4a_0a_2-a_1^2)),&\!\!\text{if~}\rho\neq (4a_2)^{-1}(4a_0a_2-a_1^2)\text{~and~}\rho\neq a_2\beta^2+a_1\beta+a_0.\\
		\end{cases}
		\end{aligned}
		\end{align}
	}
\end{lemma}

{\bf Proof.}~ By $(\ref{NDS})$, we have\begin{align}\label{l3}
\begin{aligned}
|N_{\mathbb{F}_{p^m}\backslash\{\beta\}}(a_2,a_1,a_0,\rho)|=&\frac{1}{p^m}\sum_{x\in\mathbb{F}_{p^m}\backslash\{\beta\}}\sum_{z\in\mathbb{F}_{p^m}}\zeta_p^{\mathrm{Tr}\big(z\big(\mathrm{Tr}_m^{3m}((v_2x^2+v_1x+v_0)a)-\rho\big)\big)}\\
=&|N_{\mathbb{F}_{p^m}}(a_2,a_1,a_0,\rho)|-\frac{1}{p^m}\sum_{z\in\mathbb{F}_{p^m}}\zeta_p^{\mathrm{Tr}\big((a_2\beta^2+a_1\beta+a_0-\rho)z\big)}\\
=&\begin{cases}
|N_{\mathbb{F}_{p^m}}(a_2,a_1,a_0,\rho)|-1,\quad& \text{if}~\rho= a_2\beta^2+a_1\beta+a_0;\\
|N_{\mathbb{F}_{p^m}}(a_2,a_1,a_0,\rho)|,\quad& \text{if}~\rho\neq a_2\beta^2+a_1\beta+a_0.
\end{cases}
\end{aligned}
\end{align}	

Now, for $p=2$, if $\rho= a_2\beta^2+a_1\beta+a_0$, then
\begin{align*}
\mathrm{Tr}(a_2a_1^{-2}(\rho-a_0))=\mathrm{Tr}(a_2a_1^{-2}(a_2\beta^2+a_1\beta))=2\mathrm{Tr}(a_2a_1^{-1}\beta)=0,
\end{align*}
thus by $(\ref{l1})$ and $(\ref{l3})$, we can get $(\ref{23})$.

 For odd $p$, by (\ref{l2}) and (\ref{l3}), we obtain $(\ref{p3})$ directly. 
$\hfill\Box$\\

The following lemma is useful for determining the value distribution of $|N_{\mathbb{F}_{p^m}\backslash\{\beta\}}(a_2,a_1,a_0,\rho)|$.
\begin{lemma}\label{Eq2}
	For any integer $m$ and odd prime $p$, let $\beta,\gamma_1,\gamma_0\in\mathbb{F}_{p^m}$  and $\gamma_2\in\mathbb{F}_{p^m}^{*}$, denote
	{\small	\begin{align*}
		M(\beta,\gamma_2,\gamma_1,\gamma_0)=\big\{(a_0,a_1,a_2)\in\mathbb{F}_{p^m}^3~\!|\!~a_2=\gamma_2,~(4a_2)^{-1}(4a_2 a_0-a_1^{2})=\gamma_1,~a_2\beta^2 +a_1\beta+a_0=\gamma_0\big\},
		\end{align*}}
	then
	\begin{align}\label{M}
	|M(\beta,\gamma_2,\gamma_1,\gamma_0)|=\begin{cases}
	1,\quad &\text{~if~}\gamma_0-\gamma_1=0;\\
	0,\quad &\text{~if~}\eta(\gamma_2)\eta(\gamma_0-\gamma_1)=-1;\\
	2,\quad &\text{~if~}\eta(\gamma_2)\eta(\gamma_0-\gamma_1)=1.
	\end{cases}
	\end{align}
\end{lemma}

{\bf Proof}. For $(a_0,a_1,a_2)\in\mathbb{F}_{p^m}^3$, the following system of equations
\begin{align*}
\begin{cases}
a_2=\gamma_2,\\
(4a_2)^{-1}(4a_2a_0-a_1^{2})=\gamma_1,\\
\gamma_2 \beta^2+\beta a_1+a_0=\gamma_0,
\end{cases}\end{align*}is equivalent to the system\begin{align*}
\begin{cases}
a_2=\gamma_2,\\
4\gamma_2(\gamma_0-a_2 \beta^2-\beta a_1)-a_1^{2}=4\gamma_2\gamma_1,\\
\gamma_2 \beta^2+\beta a_1+a_0=\gamma_0.
\end{cases}\end{align*}Namely,
\begin{align*}
\begin{cases}
a_2=\gamma_2,\\
(a_1-2\gamma_2 \beta)^2 =4\gamma_2(\gamma_0-\gamma_1),\\
a_0=\gamma_0-\gamma_2 \beta^2-\beta a_1.
\end{cases}
\end{align*}

Thus $(\ref{M})$ holds. $\hfill\Box$

\subsection{The Proofs for Theorems \ref{c30}-\ref{c32}}


{\bf The proof for Theorem \ref{c30}}.

By taking $D=(\alpha_1,\ldots,\alpha_n)$ and $k=2$ in $(\ref{NDS})$, we have{\small
 \begin{align}
 \begin{aligned}
 |{N}_{D}(a_{1},a_{0},\rho)|
 =&\begin{cases}
 \frac{1}{p^m}\sum\limits_{x\in D}\sum\limits_{z\in\mathbb{F}_{p^m}}\zeta_p^{\mathrm{Tr}(z(a_0-\rho))},\quad&\text{~if~} a_1=0;\\
 \frac{1}{p^m}\sum\limits_{x\in D}\sum\limits_{z\in\mathbb{F}_{p^m}}\zeta_p^{\mathrm{Tr}(z(a_1x+a_0-\rho))},\quad&\text{~if~} a_1\neq 0,
 \end{cases}\\
  =&\begin{cases}
 n,\quad&\text{~if~} a_1=0\text{~and~}\rho=a_0;\\
 0,\quad&\text{~if~} a_1=0\text{~and~}\rho\neq a_0;\\
 1,\quad&\text{~if~} \rho\in \{a_0+a_1\alpha_i~|~i=1,\ldots,n\};\\
 0,\quad&\text{~if~} \rho\notin \{a_0+a_1\alpha_i~|~i=1,\ldots,n\}.
 \end{cases}
 \end{aligned}
\end{align}}

Now for the value distribution of $|{N}_{D}(a_{1},a_{0},\rho)|$, we have the following two cases.

{\bf Case 1.} For  $\gamma_0\in\mathbb{F}_{p^m}$, if $a_1=0$ and $a_0=\gamma_0$, then
\begin{align*}
	|{N}_{D}(0,\gamma_0,\rho)|=\begin{cases}
	n,\quad & \rho=\gamma_0;\\
	0,\quad &\rho\neq \gamma_0,
	\end{cases}
\end{align*}
and the frequency is $1$.

{\bf Case 2.} For $\gamma_1\in\mathbb{F}_{p^m}^{*}$ and $\gamma_0\in\mathbb{F}_{p^m}$, if $a_1=\gamma_1$ and $a_0=\gamma_0$, then
\begin{align*}
|{N}_{D}(\gamma_{1},\gamma_0,\rho)|=\begin{cases}
1,\quad &\rho\in \{\gamma_0+\gamma_1\alpha_i~|~i=1,\ldots,n\};\\
0,\quad &\rho\notin \{\gamma_0+\gamma_1\alpha_i~|~i=1,\ldots,n\},
\end{cases}
\end{align*}
and the frequency is $1$.

By {\bf Cases} $1$-$2$, we obtain $(\ref{l31})$-$(\ref{l32})$ directly. $\hfill\Box$\\

{\bf The proof for Theorem \ref{c31}}. 

By Lemma \ref{lm3qw}, we give the complete weight enumerators for  $\mathcal{RS}_3(\boldsymbol{\alpha})$ and $\mathcal{ERS}_3(\boldsymbol{\alpha})$ depending on $p=2$ or not.

$(1)$ For $p=2$, we give the value distribution of $|{N}_{D}(a_2,a_{1},a_{0},\rho)|$ by the following four cases.

{\bf Case 1.} For $\gamma_0\in\mathbb{F}_{2^m}$, if $a_2=a_1=0$ and $a_0=\gamma_0$, then
\begin{align*}
|N_{\mathbb{F}_{2^m}}(a_2,a_1,a_0, \rho)|=\begin{cases}
2^m &\rho=\gamma_0;\\
0,\quad &\rho \neq \gamma_0,
\end{cases}
\end{align*}
and the frequency is $1$.

{\bf Case 2.} For $\gamma_1\in\mathbb{F}_{2^m}^{*}$, if $a_2=0$ and $a_1=\gamma_1$, then
\begin{align*}
|N_{\mathbb{F}_{2^m}}(a_2,a_1,a_0, \rho)|=1\quad \text{for any~}\rho\in\mathbb{F}_{2^m},
\end{align*}
and	the frequency is $2^m$.

{\bf Case 3.} For  $\gamma_2\in\mathbb{F}_{2^m}^*$, if $a_2=\gamma_2$ and $a_1=0$, then
\begin{align*}
|N_{\mathbb{F}_{2^m}}(a_2,a_1,a_0, \rho)|=1\quad \text{for any~}\rho\in\mathbb{F}_{2^m},
\end{align*}
and the frequency is $2^m$.

{\bf Case 4.} For  $\gamma_0\in\mathbb{F}_{2^m}$ and $\gamma_1,\gamma_2\in\mathbb{F}_{2^m}^*$, if $(a_2,a_1,a_0)=(\gamma_2,\gamma_1,\gamma_0)$, then

\begin{align}
\begin{aligned}
|N_{\mathbb{F}_{2^m}}(a_2,a_1,a_0,\rho)|=\begin{cases}
2,\quad &\text{~if~}\mathrm{Tr}(\gamma_2\gamma_1^{-2}(\rho-\gamma_0))=0;\\
0,\quad &\text{~if~}\mathrm{Tr}(\gamma_2\gamma_1^{-2}(\rho-\gamma_0))=1,
\end{cases}
\end{aligned}
\end{align}
and the frequency is $1$.

By the above {\bf Cases} $1$-$4$, we have
{\small \begin{align}\label{2rs3}\begin{aligned}	
\mathrm{W}(\mathcal{RS}_3(\boldsymbol{\alpha}))=&\sum_{\rho\in\mathbb{F}_{2^m}}w_\rho^{2^{m}}+2^m\sum_{\gamma_{1}\in\mathbb{F}_{2^m}^*}\prod_{\rho\in \mathbb{F}_{2^m}}w_{\rho}
	+2^m\sum_{\gamma_2\in\mathbb{F}_{p^m}^*}\prod_{\rho\in\mathbb{F}_{p^m}}w_{\rho}	\\&+\sum_{\gamma_2\in\mathbb{F}_{2^m}^*}\sum_{\gamma_1\in\mathbb{F}_{2^m}^*}\sum_{\gamma_0\in\mathbb{F}_{2^m}}\prod_{\substack{\rho\in\mathbb{F}_{2^m}\\\mathrm{Tr}(\gamma_2\gamma_1^{-2}(\rho-\gamma_0))=0}}w_{\rho}^2\\
	=&\sum_{\rho\in\mathbb{F}_{2^m}}w_\rho^{2^{m}}+(2^m-1)2^m\prod_{\rho\in \mathbb{F}_{2^m}}w_{\rho}
	+(2^m-1)2^m\prod_{\rho\in \mathbb{F}_{2^m}}w_{\rho}	\\&+(2^m-1)\sum_{\gamma_1\in\mathbb{F}_{2^m}^*}\sum_{\gamma_0\in\mathbb{F}_{2^m}}\prod_{\substack{\rho\in\mathbb{F}_{2^m}\\\mathrm{Tr}(\gamma_1^{-1}(\rho-\gamma_0))=0}}w_{\rho}^2\\
	=&\sum_{\rho\in\mathbb{F}_{2^m}}w_\rho^{2^{m}}+(2^m-1)2^{m+1}\prod_{\rho\in \mathbb{F}_{2^m}}w_{\rho}	+(2^m-1)\sum_{\gamma_1\in\mathbb{F}_{2^m}^*}\sum_{\gamma_0\in\mathbb{F}_{2^m}}\prod_{\substack{\rho\in\mathbb{F}_{2^m}\\\mathrm{Tr}(\rho)=0}}w_{\gamma_1\rho+\gamma_0}^2,
	\end{aligned}
	\end{align}}and
{\small \begin{align}\label{2ers3}\begin{aligned}	
\mathrm{W}(\mathcal{ERS}_3(\boldsymbol{\alpha}))=&\sum_{\rho\in\mathbb{F}_{2^m}}w_0w_\rho^{2^{m}}+2^m\sum_{\gamma_{1}\in\mathbb{F}_{2^m}^*}w_0\prod_{\rho\in \mathbb{F}_{2^m}}w_{\rho}
	+2^m\sum_{\gamma_2\in\mathbb{F}_{p^m}^*}w_{\gamma_2} \prod_{\rho\in\mathbb{F}_{p^m}}w_{\rho}	\\&+\sum_{\gamma_2\in\mathbb{F}_{2^m}^*}\sum_{\gamma_1\in\mathbb{F}_{2^m}^*}\sum_{\gamma_0\in\mathbb{F}_{2^m}}w_{\gamma_2}\prod_{\substack{\rho\in\mathbb{F}_{2^m}\\\mathrm{Tr}(\gamma_2\gamma_1^{-2}(\rho-\gamma_0))=0}}w_{\rho}^2\\
	=&\sum_{\rho\in\mathbb{F}_{2^m}}w_0w_\rho^{2^{m}}+(2^m-1)2^mw_0\prod_{\rho\in \mathbb{F}_{2^m}}w_{\rho}
	+2^m\sum_{\gamma_2\in\mathbb{F}_{p^m}^*}w_{\gamma_2} \prod_{\rho\in\mathbb{F}_{p^m}}w_{\rho}	\\&+\sum_{\gamma_2\in\mathbb{F}_{2^m}^*}\sum_{\gamma_1\in\mathbb{F}_{2^m}^*}\sum_{\gamma_0\in\mathbb{F}_{2^m}}w_{\gamma_2}\prod_{\substack{\rho\in\mathbb{F}_{2^m}\\\mathrm{Tr}(\rho)=0}}w_{\gamma_1\rho+\gamma_0}^2.
	\end{aligned}
	\end{align}}

$(2)$ For odd $p$,  we give the value distribution of $|{N}_{D}(a_2,a_{1},a_{0},\rho)|$ by the following three cases.

{\bf Case 1.} For $\gamma\in\mathbb{F}_{p^m}$, if $a_2=a_1=0$ and $a_0=\gamma$, then
\begin{align*}
|N_{\mathbb{F}_{p^m}}(a_2,a_1,a_0, \rho)|=\begin{cases}
p^m &\rho=\gamma;\\
0,\quad &\rho \neq \gamma,
\end{cases}
\end{align*}
and the frequency is $1$.

{\bf Case 2.} If $a_2=0$ and $a_1\neq0$, then
\begin{align*}
|N_{\mathbb{F}_{p^m}}(a_2,a_1,a_0, \rho)|=1\quad \text{for any~}\rho\in\mathbb{F}_{p^m},
\end{align*}
and	the frequency is $(p^m-1)p^m$.

{\bf Case 3.} For  $\gamma_1\in\mathbb{F}_{p^m}$ and  $\gamma_2 \in\mathbb{F}_{p^m}^{*}$, if $a_2=\gamma_2$ and $(4a_2)^{-1}(4a_0a_2-a_1^2)=\gamma_1$, then
\begin{align*}
|N_{\mathbb{F}_{p^m}}(a_2,a_1,a_0, \rho)|=\begin{cases}
1,\quad &\rho=\gamma_1;\\
1+\eta(\gamma_2)\eta(\rho-\gamma_1),\quad &\rho \neq \gamma_1,
\end{cases}
\end{align*}
and the frequency  \begin{align*}
&\#\left\{(a_2,a_1,a_0)\in\mathbb{F}_{p^m}^3~|~a_2=\gamma_2\text{~and~}(4a_2)^{-1}(4a_0a_2-a_1^2)=\gamma_1\right\}\\
=&\sum_{\tau\in\mathbb{F}_{p^m}}\#\left\{(a_2,a_1,a_0)\in\mathbb{F}_{p^m}^3~|~a_2=\gamma_2,~a_1=\tau \text{~and~}a_0=(4\gamma_2)^{-1}(4\gamma_1\gamma_2+\tau^2)\right\}\\
=&p^m.
\end{align*}

By the above {\bf Cases} $1$-$3$, we have
{\small \begin{align}\label{prs3}\begin{aligned}	
	&W(\mathcal{RS}_3(\boldsymbol{\alpha}))\\=&	\sum_{\rho\in\mathbb{F}_{p^m}}w_\rho^{p^{m}}+(p^m-1)p^m\prod_{\rho\in \mathbb{F}_{p^m}}w_{\rho}
	+ p^m\sum_{\gamma_2\in\mathbb{F}_{p^m}^*}\sum_{\gamma_1\in\mathbb{F}_{p^m}} w_{\gamma_1}\prod_{\rho\in\mathbb{F}_{p^m}\backslash\{\gamma_1\}}
	w_{\rho}^{1+\eta(\gamma_2)\eta(\rho-\gamma)}\\
	=&	\sum_{\rho\in\mathbb{F}_{p^m}}w_\rho^{p^{m}}+(p^m-1)p^m\prod_{\rho\in \mathbb{F}_{p^m}}w_{\rho}
	+\frac{(p^m-1)p^m}{2}\sum_{\epsilon\in\{-1,1\}}\sum_{\gamma_1\in\mathbb{F}_{p^m}}w_{\gamma_1}\prod_{\rho\in\mathbb{F}_{p^m}\backslash\{\gamma_1\}}
	w_{\rho}^{1+\epsilon\eta(\rho-\gamma_1)},
	\end{aligned}
	\end{align}}
and
{\small \begin{align}\label{pers3}\begin{aligned}	
	&W(\mathcal{ERS}_3(\boldsymbol{\alpha}))\\=&p^m	\sum_{\rho\in\mathbb{F}_{p^m}}w_0w_\rho^{p^{m}}+(p^m-1)p^mw_0\prod_{\rho\in \mathbb{F}_{p^m}}w_{\rho}+p^m\sum_{\gamma_2\in\mathbb{F}_{p^m}^*}\sum_{\gamma_1\in\mathbb{F}_{p^m}} w_{\gamma_2} w_{\gamma_1}\prod_{\rho\in\mathbb{F}_{p^m}\backslash\{\gamma_1\}}
	w_{\rho}^{1+\eta(\gamma_2)\eta(\rho-\gamma_1)}.
	\end{aligned}
	\end{align}}

So far, by $(\ref{2rs3})$-$(\ref{pers3})$, we complete the proof. $\hfill\Box$

{\bf The proof for Theorem \ref{c32}}.

In the similar proof as that for Theorem \ref{c31}, we prove Theorem \ref{c32} depending on $p=2$ or not.

$(1)$ For $p=2$,  we give the value distribution of $|{N}_{D}(a_{1},a_{0},\rho)|$ by the following four cases.

{\bf Case 1.} For $\gamma_0\in\mathbb{F}_{2^m}$, if $a_2=a_1=0$ and $a_0=\gamma_0$, then
\begin{align*}
|N_{\mathbb{F}_{2^m}\backslash\{\beta\}}(a_2,a_1,a_0, \rho)|=\begin{cases}
2^m-1 &\rho=\gamma_0;\\
0,\quad &\rho \neq \gamma_0,
\end{cases}
\end{align*}
and the frequency is $1$.

{\bf Case 2.} For $\gamma_{1}\in\mathbb{F}_{2^m}^*$ and $\gamma_0\in\mathbb{F}_{2^m}$, if $(a_2,a_1,a_0)=(0,\gamma_{1},\gamma_0)$ then
\begin{align*}
|N_{\mathbb{F}_{2^m}\backslash\{\beta\}}(a_2,a_1,a_0, \rho)|=\begin{cases}
0,\quad& \rho=\gamma_1\beta+\gamma_0;\\
1,\quad& \rho\neq \gamma_1\beta+\gamma_0,\\
\end{cases}
\end{align*}
and	the frequency is $1$.

{\bf Case 3.} For  $\gamma_2\in\mathbb{F}_{2^m}^*$ and $\gamma_0\in\mathbb{F}_{2^m}$, if $(a_2,a_1,a_0)=(\gamma_2,0,\gamma_0)$, then
\begin{align*}
|N_{\mathbb{F}_{2^m}\backslash\{\beta\}}(a_2,a_1,a_0,\rho)|=\begin{cases}
0,\quad &\text{~if~}\rho=\gamma_2\beta^2+\gamma_0;\\
1,\quad &\text{~if~}\rho\neq \gamma_2\beta^2+\gamma_0,
\end{cases}
\end{align*}
and the frequency is $1$.

{\bf Case 4.} For $\gamma_2,\gamma_1\in\mathbb{F}_{2^m}^*$ and  $\gamma_0\in\mathbb{F}_{2^m}$, if $(a_2,a_1,a_0)=(\gamma_2,\gamma_1,\gamma_0)$, then

\begin{align*}
|N_{\mathbb{F}_{2^m}\backslash\{\beta\}}(a_2,a_1,a_0,\rho)|=\begin{cases}
1,\quad&\text{~if~}\rho=\gamma_2\beta^2+\gamma_1\beta+\gamma_0;\\
2,\quad &\text{~if~}\mathrm{Tr}(\gamma_2\gamma_1^{-2}(\rho-\gamma_0))=0\text{~and~}\rho\neq \gamma_2\beta^2+\gamma_1\beta+\gamma_0;\\
0,\quad &\text{~if~}\mathrm{Tr}(\gamma_2\gamma_1^{-2}(\rho-\gamma_0))=1,
\end{cases}
\end{align*}
and the frequency is $1$.

By the above {\bf Cases} $1$-$4$, we have
{\small \begin{align}\label{2Ers3}\begin{aligned}	
	&\mathrm{W}(\mathcal{RS}_3(\boldsymbol{\alpha}))\\
	=&\sum_{\rho\in\mathbb{F}_{2^m}}w_\rho^{2^{m}-1}+\sum_{\gamma_{1}\in\mathbb{F}_{2^m}^*}\sum_{\gamma_{0}\in\mathbb{F}_{2^m}}\prod_{\rho\in \mathbb{F}_{2^m}\backslash\{\gamma_1\beta+\gamma_0\}}w_{\rho}
	+\sum_{\gamma_2\in\mathbb{F}_{p^m}^*}\sum_{\gamma_{0}\in\mathbb{F}_{2^m}}\prod_{\rho\in\mathbb{F}_{2^m}\backslash\{\gamma_2\beta^2+\gamma_0\}}w_{\rho}	\\&+\sum_{\gamma_2\in\mathbb{F}_{2^m}^*}\sum_{\gamma_1\in\mathbb{F}_{2^m}^*}\sum_{\gamma_0\in\mathbb{F}_{2^m}}w_{\gamma_2\beta^2+\gamma_1\beta+\gamma_0}\prod_{\substack{\rho\in\mathbb{F}_{2^m}\backslash\{\gamma_2\beta^2+\gamma_1\beta+\gamma_0\}\\\mathrm{Tr}(\gamma_2\gamma_1^{-2}(\rho-\gamma_0))=0 }}w_{\rho}^2\\
	=&\sum_{\rho\in\mathbb{F}_{2^m}}w_\rho^{2^{m}-1}+(2^m-1)\sum_{\gamma_{0}\in\mathbb{F}_{2^m}}\prod_{\rho\in \mathbb{F}_{2^m}\backslash\{\gamma_0\}}w_{\rho}
	+(2^m-1)\sum_{\gamma_{0}\in\mathbb{F}_{2^m}}\prod_{\rho\in \mathbb{F}_{2^m}\backslash\{\gamma_0\}}w_{\rho}	\\&+\sum_{\gamma_2\in\mathbb{F}_{2^m}^*}\sum_{\gamma_1\in\mathbb{F}_{2^m}^*}\sum_{\gamma\in\mathbb{F}_{2^m}}w_{\gamma}\prod_{\substack{\rho\in\mathbb{F}_{2^m}\backslash\{\gamma\}\\\mathrm{Tr}(\gamma_2\gamma_1^{-2}(\rho-\gamma))=0}}w_{\rho}^2\\
	=&\sum_{\rho\in\mathbb{F}_{2^m}}w_\rho^{2^{m}}+2(2^m-1)\prod_{\rho\in \mathbb{F}_{2^m}}w_{\rho}	+(2^m-1)\sum_{\gamma_1\in\mathbb{F}_{2^m}^*}\sum_{\gamma_0\in\mathbb{F}_{2^m}}w_{\gamma_{0}}\prod_{\substack{\rho\in\mathbb{F}_{2^m}^{*}\\\mathrm{Tr}(\rho)=0}}w_{\gamma_1\rho+\gamma_0}^2,
	\end{aligned}
	\end{align}}and
{\small \begin{align}\label{2Eers3}\begin{aligned}	
	&\mathrm{W}(\mathcal{ERS}_3(\boldsymbol{\alpha}))
	\\	=&\sum_{\rho\in\mathbb{F}_{2^m}}w_0w_\rho^{2^{m}-1}+\sum_{\gamma_{1}\in\mathbb{F}_{2^m}^*}\sum_{\gamma_{0}\in\mathbb{F}_{2^m}}w_0\prod_{\rho\in \mathbb{F}_{2^m}\backslash\{\gamma_1\beta+\gamma_0\}}w_{\rho}
	+\sum_{\gamma_2\in\mathbb{F}_{p^m}^*}\sum_{\gamma_{0}\in\mathbb{F}_{2^m}}w_{\gamma_2}\prod_{\rho\in\mathbb{F}_{2^m}\backslash\{\gamma_2\beta^2+\gamma_0\}}w_{\rho}	\\&+\sum_{\gamma_2\in\mathbb{F}_{2^m}^*}\sum_{\gamma_1\in\mathbb{F}_{2^m}^*}\sum_{\gamma_0\in\mathbb{F}_{2^m}}w_{\gamma_2}w_{\gamma_2\beta^2+\gamma_1\beta+\gamma_0}\prod_{\substack{\rho\in\mathbb{F}_{2^m}\backslash\{ \gamma_2\beta^2+\gamma_1\beta+\gamma_0\}\\\mathrm{Tr}(\gamma_2\gamma_1^{-2}(\rho-\gamma_0))=0}}w_{\rho}^2\\
	=&\sum_{\rho\in\mathbb{F}_{2^m}}w_0w_\rho^{2^{m}-1}+(2^m-1)\sum_{\gamma\in\mathbb{F}_{2^m}}w_0\prod_{\rho\in \mathbb{F}_{2^m}\backslash\{\gamma\}}w_{\rho}
	+\sum_{\gamma_2\in\mathbb{F}_{p^m}^*}\sum_{\gamma\in\mathbb{F}_{2^m}}w_{\gamma_2}\prod_{\rho\in\mathbb{F}_{2^m}\backslash\{\gamma\}}w_{\rho}	\\&+\sum_{\gamma_2\in\mathbb{F}_{2^m}^*}\sum_{\gamma_1\in\mathbb{F}_{2^m}^*}\sum_{\gamma_0\in\mathbb{F}_{2^m}}w_{\gamma_2}w_{\gamma}\prod_{\substack{\rho\in\mathbb{F}_{2^m}\backslash\{\gamma\}\\\mathrm{Tr}(\gamma_2\gamma_1^{-2}(\rho-\gamma))=0}}w_{\rho}^2\\
	=&\sum_{\rho\in\mathbb{F}_{2^m}}w_0w_\rho^{2^{m}-1}+(2^m-1)\sum_{\gamma_1\in\mathbb{F}_{2^m}}w_0\prod_{\rho\in \mathbb{F}_{2^m}\backslash\{\gamma_1\}}w_{\rho}
	+\sum_{\gamma_2\in\mathbb{F}_{p^m}^*}\sum_{\gamma_1\in\mathbb{F}_{2^m}}w_{\gamma_2}\prod_{\rho\in\mathbb{F}_{2^m}\backslash\{\gamma_1\}}w_{\rho}	\\&+\sum_{\gamma_2\in\mathbb{F}_{2^m}^*}\sum_{\gamma_1\in\mathbb{F}_{2^m}^*}\sum_{\gamma_0\in\mathbb{F}_{2^m}}w_{\gamma_2}w_{\gamma_0}\prod_{\substack{\rho\in\mathbb{F}_{2^m}^{*}\\\mathrm{Tr}(\rho)=0}}w_{\gamma_{1}\rho+\gamma_0}^2.
	\end{aligned}
	\end{align}}

$(2)$ For odd $p$, we give the value distribution of $|{N}_{D}(a_{1},a_{0},\rho)|$ by the following four cases.

{\bf Case 1.} For $\gamma_0\in\mathbb{F}_{p^m}$, if $a_2=a_1=0$ and $a_0=\gamma_0$, then
\begin{align*}
|N_{\mathbb{F}_{2^m}\backslash\{\beta\}}(a_2,a_1,a_0, \rho)|=\begin{cases}
p^m-1,\quad &\rho=\gamma_0;\\
0,\quad &\rho \neq \gamma_0,
\end{cases}
\end{align*}
and the frequency is $1$.

{\bf Case 2.} For $\gamma_0\in\mathbb{F}_{p^m}$, if $a_2=0$, $a_1\neq0$ and $a_1\beta+a_0=\gamma_0$ then
\begin{align*}
|N_{\mathbb{F}_{2^m}\backslash\{\beta\}}(a_2,a_1,a_0, \rho)|=\begin{cases}
0,\quad &\rho=\gamma_0;\\
1,\quad &\rho \neq \gamma_0,
\end{cases}
\end{align*}
and the frequency is $p^m-1$.

{\bf Case 3.} For $\gamma_1\in\mathbb{F}_{p^m}$ and $\gamma_2\in\mathbb{F}_{p^m}^*$, if $a_2=\gamma_2 $ and $(4a_2)^{-1}(4a_0a_2-a_1^2)=a_2\beta^2+a_1\beta+a_0=\gamma_1$, then
\begin{align*}
|N_{\mathbb{F}_{2^m}\backslash\{\beta\}}(a_2,a_1,a_0, \rho)|=\begin{cases}
0,\quad &\rho=\gamma_1;\\
1+\eta(\gamma_2)\eta(\rho-\gamma_1),\quad &\rho \neq \gamma_1,
\end{cases}
\end{align*}
and the frequency  \begin{align*}|M(\beta,\gamma_2,\gamma_1,\gamma_1)|=1.
\end{align*}

{\bf Case 4.} For $\gamma_2\in\mathbb{F}_{p^m}^{*}$ and $\gamma_1,\gamma_0\in\mathbb{F}_{p^m}$ with $\gamma_1\neq\gamma_0$, if $a_2=\gamma_2$,  $(4a_2)^{-1}(4a_0a_2-a_1^2)=\gamma_1$ and $a_2\beta^2+a_1\beta+a_0=\gamma_0$, then
\begin{align*}
|N_{\mathbb{F}_{2^m}\backslash\{\beta\}}(a_2,a_1,a_0, \rho)|=&\begin{cases}
1,~&\text{~if~}\rho=\gamma_1\text{~and~}\rho\neq \gamma_0;\\
\eta(\gamma_2)\eta(\rho-\gamma_1),~&\text{~if~}\rho\neq \gamma_1\text{~and~}\rho= \gamma_0;\\
1+\eta(\gamma_2)\eta(\rho-\gamma_1),~&\text{~if~}\rho\neq \gamma_1\text{~and~}\rho\neq \gamma_0,
\end{cases}
\end{align*}
and the frequency  \begin{align*}|M(\beta,\gamma_2,\gamma_1,\gamma_0)|=\begin{cases}
0,\quad&\eta(\gamma_2)\eta(\gamma_0-\gamma_1)=-1;\\
2,\quad&\eta(\gamma_2)\eta(\gamma_0-\gamma_1)=1.
\end{cases}
\end{align*}

By the above {\bf Cases} $1$-$4$, we have
	{\small \begin{align}\label{prs32}\begin{aligned}	
&\mathrm{W}(\mathcal{RS}(\boldsymbol{\alpha}))\\=&	\sum_{\rho\in\mathbb{F}_{p^m}}w_\rho^{p^m-1}+(p^m-1)\sum_{\gamma_0\in\mathbb{F}_{p^m}}\prod_{\rho\in \mathbb{F}_{p^m}\backslash\{\gamma_0\}}w_{\rho}
+\sum_{\gamma_2\in\mathbb{F}_{p^m}^{*}}\sum_{\gamma_1\in\mathbb{F}_{p^m}}\prod_{\rho\in\mathbb{F}_{p^m}\backslash\{\gamma_1\}}
	w_{\rho}^{1+\eta(\gamma_2)\eta(\rho-\gamma_1)}\\
	&+2\sum_{\gamma_2\in\mathbb{F}_{p^m}^{*}}\sum_{\substack{(\gamma_1,\gamma_0)\in\mathbb{F}_{p^m}^2\\\eta(\gamma_2)\eta(\gamma_0-\gamma_1)=1}}w_{\gamma_1}w_{\gamma_0}\prod_{\rho\in\mathbb{F}_{p^m}\backslash\{\gamma_1,\gamma_0\}}w_{\rho}^{1+\eta(\gamma_2)\eta(\rho-\gamma_1)}\\
	=&	\sum_{\rho\in\mathbb{F}_{p^m}}w_\rho^{p^m-1}+(p^m-1)\sum_{\gamma\in\mathbb{F}_{p^m}}\prod_{\rho\in \mathbb{F}_{p^m}\backslash\{\gamma\}}w_{\rho}+\frac{p^m-1}{2}\sum_{\epsilon\in\{-1,1\}}\sum_{\gamma_1\in\mathbb{F}_{p^m}}\prod_{\rho\in\mathbb{F}_{p^m}\backslash\{\gamma_1\}}
	w_{\rho}^{1+\epsilon\eta(\rho-\gamma_1)}\\
	&+(p^m-1)\sum_{\epsilon\in\{-1,1\}}\sum_{\gamma_1\in\mathbb{F}_{p^m}}\sum_{\substack{\gamma_0\in\mathbb{F}_{p^m}\\\eta(\gamma_0)=\epsilon}}w_{\gamma_1}w_{\gamma_0+\gamma_{1}}\prod_{\rho\in\mathbb{F}_{p^m}\backslash\{\gamma_1,\gamma_0+\gamma_1\}}w_{\rho}^{1+\epsilon\eta(\rho-\gamma_1)},
	\end{aligned}
	\end{align}}and
	{\small \begin{align}\label{pers32}\begin{aligned}	
&\mathrm{W}(\mathcal{ERS}(\boldsymbol{\alpha}))\\=&w_0	\sum_{\rho\in\mathbb{F}_{p^m}}w_\rho^{p^m-1}+(p^m-1)\sum_{\gamma_0\in\mathbb{F}_{p^m}}w_0\prod_{\rho\in \mathbb{F}_{p^m}\backslash\{\gamma_0\}}w_{\rho}
+\sum_{\gamma_2\in\mathbb{F}_{p^m}^{*}}\sum_{\gamma_1\in\mathbb{F}_{p^m}}w_{\gamma_2}\prod_{\rho\in\mathbb{F}_{p^m}\backslash\{\gamma_1\}}
	w_{\rho}^{1+\eta(\gamma_2)\eta(\rho-\gamma_1)}\\
	&+2\sum_{\gamma_2\in\mathbb{F}_{p^m}^{*}}\sum_{\substack{(\gamma_1,\gamma_0)\in\mathbb{F}_{p^m}^2\\\eta(\gamma_2(\gamma_0-\gamma_1))=1}}w_{\gamma_2}w_{\gamma_1}w_{\gamma_0}\prod_{\rho\in\mathbb{F}_{p^m}\backslash\{\gamma_1,\gamma_0\}}w_{\rho}^{1+\eta(\gamma_2)\eta(\rho-\gamma_1)}\\
	=&w_0	\sum_{\rho\in\mathbb{F}_{p^m}}w_\rho^{p^m-1}+(p^m-1)\sum_{\gamma_0\in\mathbb{F}_{p^m}}w_0\prod_{\rho\in \mathbb{F}_{p^m}\backslash\{\gamma_0\}}w_{\rho}
	+\sum_{\gamma_2\in\mathbb{F}_{p^m}^{*}}\sum_{\gamma_1\in\mathbb{F}_{p^m}}w_{\gamma_2}\prod_{\rho\in\mathbb{F}_{p^m}\backslash\{\gamma_1\}}
	w_{\rho}^{1+\eta(\gamma_2)\eta(\rho-\gamma_1)}\\
	&+2\sum_{\epsilon\in\{-1,1\}}\sum_{\substack{\gamma_2\in\mathbb{F}_{p^m}^{*}\\\eta(\gamma_2)=\epsilon}}\sum_{\gamma_1\in\mathbb{F}_{p^m}}\sum_{\substack{\gamma_0\in\mathbb{F}_{p^m}\\\eta(\gamma_0)=\epsilon}}w_{\gamma_2}w_{\gamma_1}w_{\gamma_0+\gamma_{1}}\prod_{\rho\in\mathbb{F}_{p^m}\backslash\{\gamma_1,\gamma_0+\gamma_1\}}w_{\rho}^{1+\epsilon\eta(\rho-\gamma_1)}.
	\end{aligned}
	\end{align}}

So far, by $(\ref{2Ers3})$-$(\ref{pers32})$, we complete the proof. $\hfill\Box$

\end{document}